\documentclass[epj]{svjour}

\usepackage{graphicx,color}
\usepackage{amsmath}
\usepackage{amssymb}
\usepackage{bm}
\usepackage{epstopdf}
\usepackage{color}

\newcommand{\slr}{$(T_1T)^{-1}$}

\newcommand{\sro}{Sr$_2$RuO$_4$}

\begin{document}

\title{$^{75}$As NMR-NQR study in superconducting LiFeAs}

\author{Seung-Ho Baek\inst{1} \fnmsep\thanks{\email{sbaek.fu@gmail.com}} \and Hans-Joachim Grafe\inst{1} \and Franziska Hammerath\inst{1} \and Madeleine
Fuchs\inst{1} \and Christian Rudisch\inst{1} \and Luminita
Harnagea\inst{1} \and Saicharan Aswartham\inst{1} \and Sabine
Wurmehl\inst{1} \and Jeroen {van den Brink}\inst{1} \and Bernd
B\"{u}chner\inst{1}}

\institute{Leibniz Institute for Solid State and Materials
Research IFW-Dresden, PF 270116, 01171 Dresden, Germany}

\date{Received: date / Revised version: date}

\abstract{We report results of $^{75}$As nuclear magnetic
resonance (NMR) and nuclear quadrupole resonance (NQR) experiments
as well as $^7$Li NMR on different samples of self flux grown
LiFeAs and 5\% Co doped LiFeAs single crystals, and a
polycrystalline LiFeAs sample. We were able to distinguish the
samples by their slightly different quadrupole frequencies,
$\nu_Q$, which is a direct measure of the electric field gradient
(EFG) at the As site. Interestingly, samples with a large
quadrupole frequency appear to show a different Knight shift and
spin lattice relaxation in the superconducting state from those
with a lower $\nu_Q$, yet all the samples are clearly
superconducting. For sample S1 which has the largest $\nu_Q$, we
find constant Knight shift $\mathcal{K}$ across $T_c$ for a
certain direction of the magnetic field and a peculiar upturn of
the NQR spin lattice relaxation rate \slr\ below $T_c$. In contrast,
samples with a lower $\nu_Q$ exhibit the expected behavior for a
singlet superconductor: a drop of $\mathcal{K}$ and \slr\ for both NMR 
and NQR below $T_c$. Our results show that already tiny changes in stoichiometry
uncovered by slightly different NQR frequencies lead to very
different behavior of the NMR parameters in the superconducting
state of LiFeAs. Different possibilities will be discussed which
may explain the contrasting behavior.}

\maketitle


\section{Introduction}
\label{intro}

Among the recently discovered superconducting iron pnictides
\cite{kamihara08,rotter08}, LiFeAs is a rare member exhibiting
superconductivity with $T_c\sim18$ K in a stoichiometric form
without dopants or pressure. Yet, the Li content in this compound
is difficult to control and seems to have a large influence on the physical 
properties. For example, Li deficiencies of only $\sim 1$\% 
greatly suppress superconductivity \cite{pitcher10}. Probably related to the 
sensitivity of the Li concentration, the superconducting ground
state of LiFeAs and the pairing mechanism are still under debate.

Small angle neutron scattering (SANS) and
angle resolved photoemission spectroscopy (ARPES)
\cite{inosov10a,borisenko10,kordyuk11} suggest that LiFeAs could be a
weakly electron-phonon coupled conventional-type superconductor. However, a recent Raman 
scattering study did not find evidence for substantial electron-phonon-coupling and no
superconductivity-induced phonon anomalies \cite{um12}.
Furthermore, while the spin density wave (SDW) state is absent in LiFeAs, there is
evidence that weak local moments \cite{zhang09} and magnetic
fluctuations are still present in the normal state, putting LiFeAs
close to a magnetic instability \cite{pratt09b,qureshi12}.  Theoretical analyses
of the electronic band-structure find a superconducting order
parameter of $s_\pm$ type driven by collinear antiferromagnetic
fluctuations \cite{platt11}. The fully gapped $s_{\pm}$ superconductivity  
is also suggested by heat transport \cite{tanatar11}, 
magnetic penetration depth measurements \cite{hashimoto12}, and 
inelastic neutron scattering (INS) study \cite{taylor11}.
On the other hand, Brydon et al. \cite{brydon11} proposed spin-triplet
$p$-wave pairing in this system. Being in agreement with the theoretical argument, vortex 
properties \cite{pramanik11} and $H_{c2}$ measurements \cite{cho11} for $H
\parallel ab$ show a very similar behavior as in the supposedly
triplet superconductor \sro, and quasi particle interference (QPI)
patterns measured by scanning tunneling microscopy (STM)
\cite{haenke12} supports an elementary $p$-wave symmetry, rather than singlet pairing
symmetries ($s^{\pm}$- or $d$-wave).

The so far reported NMR measurements of LiFeAs powder samples
indicate the existence of magnetic correlations from the analysis
of the Korringa relation \cite{jeglic10,li10}. In the
superconducting state, the Knight shift shows a sharp drop at
$T_c$ which is suggestive of spin-singlet superconductivity. In
the only NMR study on single crystals Ma \textit{et al.} argue
that the absence of the spin density wave ordering in LiFeAs is
due to off-stoichiometry and/or lattice defects, and find two
different Li sites in their superconducting samples
\cite{ma10}.

Here, we report detailed NMR and NQR results on three single
crystal LiFeAs samples, as well as on a 5\% Co doped single
crystal, and a polycrystalline LiFeAs sample. We show that the NQR
frequency is highly sensitive to the Li content, and that samples
with tiny differences in stoichiometry can be distinguished by
their different quadrupole frequencies. A doping dependence of the
quadrupole frequency has already been found in other iron
pnictides \cite{lang10}, and similar changes of $\nu_Q$ with
Co doping in LiFeAs indicate that also Li deficiencies change the
doping level of the samples. Interestingly, samples with a large
quadrupole frequency appear to show a different Knight shift and
spin lattice relaxation in the superconducting state from those
with a lower $\nu_Q$. Sample S1 which has the largest $\nu_Q$,
exhibits a constant Knight shift $\mathcal{K}$ across $T_c$ for a
certain direction of the magnetic field and a peculiar upturn of
the spin lattice relaxation rate \slr\ below $T_c$. In contrast,
samples with a lower $\nu_Q$ exhibit a drop of $\mathcal{K}$ and \slr\ below
$T_c$. Our results show that already tiny changes in stoichiometry
uncovered by slightly different NQR frequencies lead to very
different behavior of the NMR parameters in the superconducting
state of LiFeAs. The tiny differences in stoichiometry and
their large impact on the superconducting properties may also
account for the contradicting results reported so far.

\section{Sample preparation and Experimental details}
\label{sampleprep}

The LiFeAs single crystals were grown by a self-flux method using
a molar ratio of Li:Fe:As = 3:2:3 similar to
ref.~\cite{morozov10}.
The stoichiometry of the samples has been checked by inductively
coupled plasma mass spectroscopy (ICPMS). 20 mg of the LiFeAs
single crystal were dissolved in a leak free glass ampoule in nitric acid. The 
molar ratio Li:Fe:As is found to be 
0.99:1.00:1.00, consistent with a stoichiometric LiFeAs
composition \cite{morozov10}. ARPES measurements performed on a
LiFeAs single crystal from the same batch were found to be in
agreement with an exact stoichiometry \cite{borisenko10}.

The susceptibility of our samples exhibits a sharp superconducting
transition at $T_c \sim 18$ K \cite{morozov10,inosov10a}. A clear
anomaly at $T_c$ in the specific heat measurement
\cite{stockert11}, and the large residual resistance ratio (RRR)
\cite{morozov10,heyer11}, with the lowest residual resistivity
observed so far in iron pnictides \cite{tapp08,wang08,song10}
indicate that the samples are clean superconductors. Furthermore,
STM measurements yield a defect concentration $< 1$ \%
\cite{haenke12}. Another method to check the sample quality is the
measure of the linewidth of a NQR spectrum or, equivalently, of
the satellite transitions ($I_z = -3/2 \leftrightarrow -1/2$ and
$I_z = +1/2 \leftrightarrow +3/2$) of a NMR spectrum, because
those measure directly the distribution of the electric field
gradient (EFG) which is caused by disorder or defects in the
sample, while the broadening of the central transition ($I_z =
-1/2 \leftrightarrow +1/2$) is dominantly magnetic in origin. As
we will show below, all samples exhibit very narrow linewidth of
both $^{75}$As NQR lines and $^{7}$Li NMR satellites which prove
the absence of large amounts of defects in our samples and a
stoichiometry of close to 1:1:1.

$^{75}$As (nuclear spin $I=3/2$) NMR/NQR, and $^{7}$Li ($I=3/2$)
NMR measurements were carried out in five different LiFeAs samples: four
single crystalline samples: S1, S2, S3, and Co 5\%-doped LiFeAs, and a
polycrystalline sample, where the single crystals S1, S2, and S3 are from
the same batch. Due to the extreme sensitivity of the samples to
air and moisture, all the samples were carefully sealed into
quartz tubes filled with Ar gas. It could happen that thermal cycling
damages the sealing of the quartz tubes that contain the
sample, so that, if the sample probe has to be taken out of the cryostat, the
sample is easily degraded by contact with
air. For this reason we could not obtain full data
sets for all of the samples since the NMR and NQR signals become negligibly
weak in a sample which had contact with air.

Since the local symmetry at the $^{75}$As is axial (tetragonal),
the nuclear quadrupole frequency $\nu_Q$ can be determined
directly from the resonance frequency of the NQR spectrum for the
$^{75}$As and by the splitting of the satellites for the $^{7}$Li.
The Knight shift was obtained by measuring the central transition. The nuclear
spin-lattice relaxation rates $T_1^{-1}$ were measured by
saturation and inversion recovery methods. For exact orientation
of the crystals with respect to the static magnetic field a single
axis goniometer has been used for most of the measurements.

\section{Experimental Results}

\subsection{$^{75}$As NQR}

For the $^{75}$As in a non-cubic environment, the four-fold
degeneracy of the nuclear spin of $I=3/2$ is partially lifted by
the interaction between the nuclear quadrupole moment $Q$ and the
surrounding EFG, $eq$, which allows a NQR resonance at a frequency
given by $\nu_Q\equiv e^2qQ/h$ where $h$ is Planck's constant and
$e$ the electron charge. We find a very narrow NQR line near 21.5
MHz at room temperature, as shown in the inset of Fig. 1. The
$^{75}$As NQR spectra have a width ranging from 60 kHz (S2) to 80
kHz (S1) at room temperature, which is a factor of 2-3 narrower
than 170 kHz reported earlier in powder samples of LiFeAs
\cite{li10}, and even significantly smaller than the values
reported for other undoped (non-superconducting) iron pnictides:
100 kHz from NMR satellites in BaFe$_2$As$_2$ \cite{kitagawa08},
308 kHz and 385 kHz in La- and SmOFeAs, respectively
\cite{lang10}, and 480 kHz in CaFe$_2$As$_2$ \cite{curro09}. This
is even more surprising regarding the high quadrupole frequency of
LiFeAs with respect to the other iron pnictides, and indicates the
high homogeneity of all of our single crystals.

The temperature dependences of the nuclear quadrupole frequency
$\nu_Q$ of the $^{75}$As are shown in Fig. 1. The results obtained
in the polycrystalline sample and the 5\% Co doped single crystal
sample are also compared. We find that $\nu_Q$, or the EFG $eq$,
varies with the samples investigated. Despite the different
$\nu_Q$ at a given temperature, the temperature dependences of
$\nu_Q$ of all the samples are very similar among one another,
i.e., $\nu_Q$ decreases with decreasing $T$, being saturated at
low temperatures. Such a strong temperature dependence of $\nu_Q$
is commonly observed in iron pnictides \cite{kitagawa08,baek09},
and is attributed to mostly electronic effects due to complicated
multi-orbitals of the iron ion ($3d^6$), since the lattice
(thermal vibration) contribution, usually, leads to an opposite
(and very weak) temperature dependence. However, we can
distinguish the samples clearly by their different values at a
given temperature, which resembles a doping dependence similar to
other pnictides \cite{lang10,baek11} even though samples S1-S3 are
from the same batch, and should have the same composition
according to ICPMP and ARPES \cite{morozov10,borisenko10}.
Interestingly, we observe that $\nu_Q(T)$ of the polycrystalline
sample is very similar to that of S1, while $\nu_Q(T)$ of the
Co-doped sample is even below those of S2 and S3. It appears that samples with
a larger $\nu_Q$ reveal a peculiar temperature dependence of the
relaxation rates and/or the Knight shift in the superconducting
state, as shown below, whereas samples with a lower $\nu_Q$ show
more normal behaviors, as expected in typical spin-singlet
superconductors. We note that $\nu_Q(T)$ of the powder samples
obtained by Li et al. \cite{li10} appears to be close to our
Co-5\% doped single crystal.

\begin{figure}
\centering
\includegraphics[width=\linewidth]{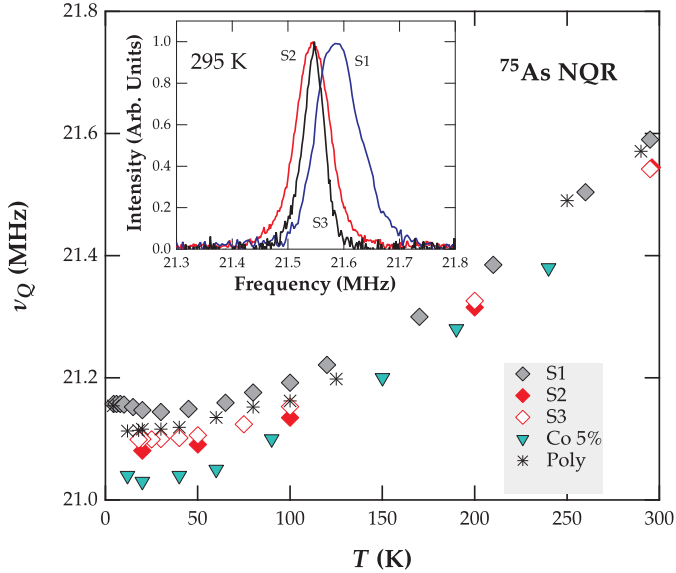}
\caption{The quadrupole frequency, $\nu_Q$, decreases with
decreasing temperature and almost saturates at low temperatures.
While the temperature dependences of $\nu_Q$ of all the samples
are similar, the values of the sample S1 are noticeably larger
than those of the samples S2 and S3, as clearly shown in the inset. For
comparison, $\nu_Q(T)$ for polycrystalline and Co-doped samples
are shown, too. Note that the linewidth is also strongly 
sample-dependent, yielding 80, 60, and 40 kHz for S1, S2, and S3, 
respectively at room temperature.}
\end{figure}

Fig. 2 shows the $^{75}$As NQR measurement of the spin lattice
relaxation rate divided by temperature, \slr. For all samples,
\slr\ slightly decreases with decreasing $T$ and approaches a
constant value below $\sim 150$ K down to $T_c$. Below $T_c$,
however, the temperature dependence of \slr\ varies among the
samples. Due to the opening of the superconducting gap, the
electron density of states at the Fermi level to which \slr\ is
proportional, is reduced rapidly, and thus \slr\ is expected to
decrease accordingly. While S3 exhibits such a rapid drop at
$T_c$, S1 does not. Nevertheless, the fact that the upturn occurs
at $T_c$ suggests that the unusual behavior is associated with
superconductivity, as will be discussed below in comparison to the
NMR \slr. For the Co-doped sample, the enhancement of \slr\ is
suppressed, while maintaining the constant value from above $T_c$.
The polycrystalline sample also shows a strong enhancement of
\slr\ below $T_c$, similar to that of S1. In all cases,
superconductivity has been confirmed at the same time by a strong
change in the resonance frequency of the NMR circuit, which is
proportional to the ac susceptibility of the sample. We note that
both S1 and the polycrystalline sample which reveal a similar
behavior of \slr\ possess larger $\nu_Q$ values than the other
samples, as shown in Fig. 1. Tentatively, this trend may suggest
that the strength of the EFG has a relationship with the
underlying physics which may cause the different behaviors of
\slr\ in the superconducting state.

\begin{figure}
\centering
\includegraphics[width=\linewidth]{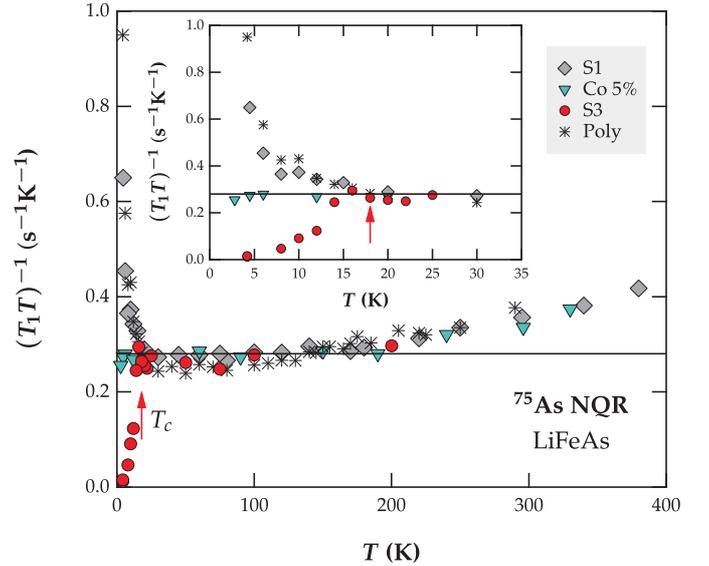}
\caption{$^{75}$As NQR spin-lattice relaxation rates divided by
$T$, \slr, as a function of $T$ in zero field. While \slr\ of all
four samples are essentially the same above $T_c$, those change
drastically below $T_c$.  For both the single crystal S1 and the
polycrystalline sample, \slr\ rises below $T_c$. 5\% Co-doping
suppresses the enhancement below $T_c$, but still without a
decrease. On the contrary, data from another single crystal S3
reveals a rapid drop, as expected in the superconducting state.}
\end{figure}

\subsection{$^{7}$Li NMR spectra}

Fig. 3 shows $^{7}$Li NMR spectra including the two satellites
obtained for the crystal S3. At 200 K, the spectra for both
directions along $c$ and $ab$ reveal very sharp lines which allow
us to determine $\nu_Q=33$ kHz accurately. This value is also
consistent with the value of $\nu_Q=34$ kHz estimated by Jegli\v{c} et
al. \cite{jeglic10} by an echo decay measurement. The almost
identical linewidth of cental and satellite lines indicates that
the broadening mechanism is mainly magnetic, and the distribution
of the EFG is negligibly small. At low temperatures, the $^{7}$Li
spectra broaden significantly, unlike the $^{75}$As NMR-NQR
spectrum whose linewidth increases only for S1 at low
temperatures. This indicates that the $^{7}$Li nuclei experience
quasi-static spin fluctuations at low temperatures, particularly
for $H$ along the $c$ axis.

\begin{figure}
\centering
\includegraphics[width=\linewidth]{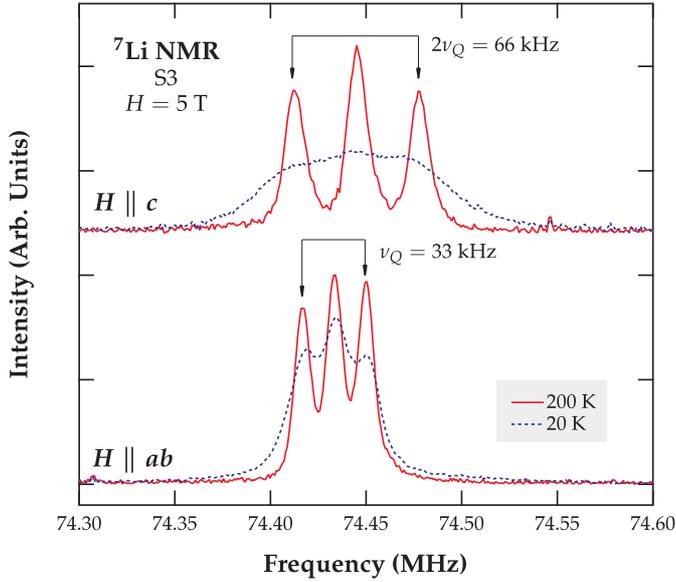}
\caption{$^7$Li NMR spectra at 200 K and 20 K at $H = 4.9994$ T,
with the $^7$Li quadrupole frequency, $\nu_Q=33$ kHz. The
separation between the satellite lines for $H\parallel c$
corresponds to $2\nu_Q$, as expected in the tetragonal symmetry
(the asymmetry parameter $\eta=0$). The linewidth, which is the
same for central and satellite lines, for $H\parallel ab$
($H\parallel c$) increases from 9 kHz (11 kHz) at 200 K to 20 kHz
(40 kHz) at 20 K.}
\end{figure}

We emphasize that our $^7$Li NMR spectra assure that there is only
one single Li site in our crystals. This is in stark contrast with
the crystals used by Ma \textit{et al.} \cite{ma10} where two
$^7$Li resonances are observed, indicating the presence of
inequivalent Li sites in their crystals. Therefore our results are
not consistent with the claim that the absence of magnetism is due
to the off-stoichiometry or defects. Rather, we argue in this
paper that the stoichiometry is a critical parameter governing the
nature of magnetism and superconductivity in LiFeAs.

\subsection{$^{75}$As NMR Knight shifts}
\label{knightshift}

Fig.~4 (a) shows the $T$-evolution of the $^{75}$As NMR central
line of the single-crystal S1 measured with $H$ parallel to the
$ab$-plane. At high temperatures, the spectrum shifts down with
decreasing $T$, and approaches a constant frequency below $\sim
50$ K. Unexpectedly, we observed that the spectrum (drawn with
thick lines below $T_c$) maintains the resonance frequency in the
superconducting state down to 4.2 K. This behavior drastically
changes when $H$ is rotated out of the $ab$-plane by $\sim
2^\circ$ -- now the $^{75}$As NMR line shows visible shifts below
$T_c$, see Fig. 4 (b). Nevertheless, this indicates that the
sample is superconducting.

\begin{figure}
\centering
\includegraphics[width=\linewidth]{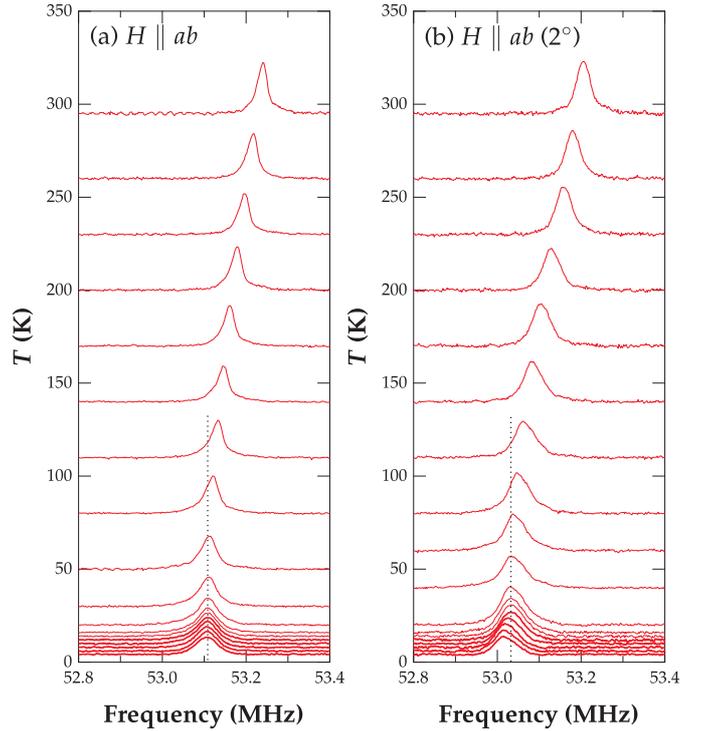}
\caption{\label{fig:As_spec}
$^{75}$As NMR spectra of the central transition ($I_z=1/2
\leftrightarrow -1/2$) at an external field of $H=7.0494$ T as a
function of temperature for the single crystal S1. Spectra in the
superconducting state are shown as thick lines. (a) Spectra for
$H\parallel ab$. Note that the resonance frequency of the spectra
does not change upon going through $T_c$. (b) Same as in (a), but
with tilting the sample by $2^\circ$. Although the temperature
dependence of the shift above $T_c$ is similar to the case of the
exact alignment, the resonance frequency drops at $T_c(H)\sim 13$
K, indicating a vanishing spin susceptibility.}
\end{figure}

Due to the large NQR frequency $\nu_Q$, there should be a
substantial shift of the central transition for $H \parallel ab$ that is strongly angle
dependent, due to the second order quadrupole effect. In an
uniaxial symmetry, which we confirmed by $^7$Li NMR spectra, the
second order quadrupole shift for a nucleus with spin  $I=3/2$ is
given by \cite{bennet}
\begin{equation}
\label{}
\Delta\nu (\theta) =
\frac{3\nu_Q^2}{16\gamma_n H}(1-\cos^2\theta)(1-9\cos^2\theta),
\end{equation}
where $\theta$ is the angle between the tetragonal $c$ axis and $H$.

$\Delta\nu (\theta)$ has been subtracted from the total shift of
the NMR lines to extract the Knight shift shown in Fig.~5. In our
case, the extreme sensitivity of $\Delta\nu$ to $\theta$ caused by
the large $\nu_Q$ is indeed a benefit since one can take advantage
of it in order to align the sample with great accuracy. In
particular, it is very useful for $H \parallel ab$, since
$\Delta\nu$ is simply the maximum when $\theta=90^\circ$.
Furthermore, we confirmed by exact diagonalization of the nuclear
Hamiltonian the validity of the second order correction.

The resulting Knight shift $\mathcal{K}$ shown in Fig.~5(a)
decreases slowly upon lowering the temperature, flattening out at
around 50 K, being in agreement with the results reported
previously \cite{li10,jeglic10}, as already reflected in the raw
data (see Fig. 4). The total Knight shift $\mathcal{K}$ consists
of a spin and an orbital contribution,
$\mathcal{K}=\mathcal{K}_\text{spin}+\mathcal{K}_\text{orb}$,
where the latter involves the orbital motion of the conduction
electrons and is usually temperature independent. Here,
$\mathcal{K}_\text{spin}$ is directly proportional to the spin
susceptibility, $\mathcal{K}_\text{spin}=A\chi_\text{spin}$, where
$A$ is the hyperfine coupling strength. Therefore, $\mathcal{K}$
should vanish at $T\ll T_c$ if the spin state of Cooper pairing is
singlet. Usually, the hyperfine coupling constant is extracted
from plots of $\mathcal{K}$ versus the bulk susceptibility,
$\chi$.
We extracted the hyperfine coupling constants for
temperatures $T>170$ K, because below the temperature the Knight shift does not scale with the
susceptibility $\chi$ due to a magnetic impurity contribution which affects
only $\chi$. For $H \parallel ab$ we obtain
$A_{ab}=6.3$ T/$\mu_B$, and for $H\parallel c$ $A_c=0.95$
T/$\mu_B$. These values appear to be
strongly anisotropic, compared to other iron pnictides, e.g., 
$A_{ab}=3.87$ T/$\mu_B$ and $A_c=2.61$ T/$\mu_B$ for isostructural NaFeAs 
\cite{kitagawa11} and $A_{ab} = 2.64$ T/$\mu_B$ and $A_c =
1.88$ T/$\mu_B$ for BaFe$_2$As$_2$ \cite{kitagawa08}.

A scaling of the Knight shifts measured for different orientations
and for different nuclei would indicate that all nuclei probe the
same component of the spin susceptibility. Such a behavior has
been found for optimally doped LaO$_{0.9}$F$_{0.1}$FeAs
\cite{grafe09}. Instead, we find that for $^{75}$As
$\mathcal{K}^{ab}$ does not give a linear relation with
$\mathcal{K}^{c}$.\footnote{We do not show $\mathcal{K}^c$ here,
because we measured our samples only with a single axis
goniometer. Therefore, while one can reach a perfect alignment for
$H \parallel ab$, there may be small deviations for $H \parallel
c$ which lead to additional second order quadrupole correction.
This affects only the absolute value of $\mathcal{K}^c$, but not
the temperature dependence which can be compared with
$\mathcal{K}^{ab}$.} Likewise, the Knight shift of the $^7$Li
nucleus does not scale with the Knight shift measured at the
$^{75}$As for both directions. Since there are several bands
crossing the Fermi level in the iron pnictides, one can expect
each band to make a different contribution to the total spin
susceptibility with different hyperfine couplings to different
bands. Such is the case for $^{17}$O NMR in Sr$_2$RuO$_4$, where
the oxygen simultaneously couples to multiple bands with different
temperature-dependent susceptibilities \cite{imai98}. The strong
angular dependence of $\mathcal{K}$ is also observed for the
$^{17}$O Knight shift in Sr$_2$RuO$_4$, whose sign changes with
angle at low temperatures as in our case. While the spin
susceptibility $\chi_\text{spin}$ is always positive, the Knight
shift can be negative since $\mathcal{K}_\text{spin}$ can be
decomposed into two components: $ \mathcal{K}_\text{spin} =
A_s\chi_s + A_\text{cp}\chi_{\text{non-}s} $ where $A_{s}$ is the
direct Fermi contact hyperfine coupling to $s$-electrons and
$A_\text{cp}$ arises from core polarization of inner $s$-shells
due to non-$s$ electrons ($p$ or $d$) \cite{abragam-nmr}. Here,
$A_s$ is always positive whereas $A_\text{cp}$ is always negative.
Thus, if $\chi_{\text{non-}s}$ is strongly angle and temperature
dependent, $\mathcal{K}$ will change accordingly, possibly
reversing its sign. Although this may catch the essential
features, the whole $T$- and angle-dependencies of our data
obtained in LiFeAs are not correctly understood quantitatively
within this simple picture. This is likely due to the multi-band
structure as mentioned above. In this case, the spin
susceptibility from each band may have quite different response to
temperature and field direction, depending on the overlap with
$p$-orbitals of the As ion, resulting in the observed angular and
temperature dependencies of the Knight shift.

Now, we focus on the temperature dependence of $\mathcal{K}$ at
low temperatures, which is shown in  Fig.~5(b). Upon going through
the superconducting $T_c$, the Knight shift $\mathcal{K}^{ab}$ for
the sample S1 does not show any change, which was already
manifested in the raw data [Fig. 4(a)].

Furthermore, in the superconducting state the behavior of
$\mathcal{K}$ changes when the field $H$ is tilted by just
2$^\circ$ out of the $ab$ plane: there is a clear drop of
$\mathcal{K}$ at $T_c$ and it approaches to a finite value as
$T\rightarrow 0$ (see Fig. 5). For an angle of 5.6$^\circ$,
$\mathcal{K}$ is small and even upturns slightly below $T_c$, but
approaches a similar value for $2^\circ$-off case, too, as
$T\rightarrow 0$. The temperature dependence of $\mathcal{K}$ for
a tilting of 2$^\circ$ is similar to the data measured in the
polycrystalline sample \cite{jeglic10,li10}, and further indicates
that our crystals are superconducting. In polycrystalline samples,
however, one may argue that whether $\mathcal{K}^{ab}$ is constant
for precisely in-plane field is very difficult to be confirmed due
to the inevitable ambiguity in assigning the singularity for
$H\parallel ab$ in a powder pattern.

However, the results measured in another single crystal S2 exhibit
totally different behavior. $\mathcal{K}$ for S2 is slightly
larger than for S1 above $T_c$, and $\mathcal{K}^{ab}$ shows a
clear drop at $T_c$. With tilting the crystal by $3^\circ$, a
similar behavior of $\mathcal{K}$ was observed, unlike the strong
angle dependence obtained for S1. Such a strong variation of
$\mathcal{K}$ in two samples grown in the same batch is
surprising, and we interpret this as an indication that LiFeAs is
located in the vicinity of an instability which may cause the
extreme sensitivity of the physical properties.

\begin{figure}
\centering
\includegraphics[width=3.5in]{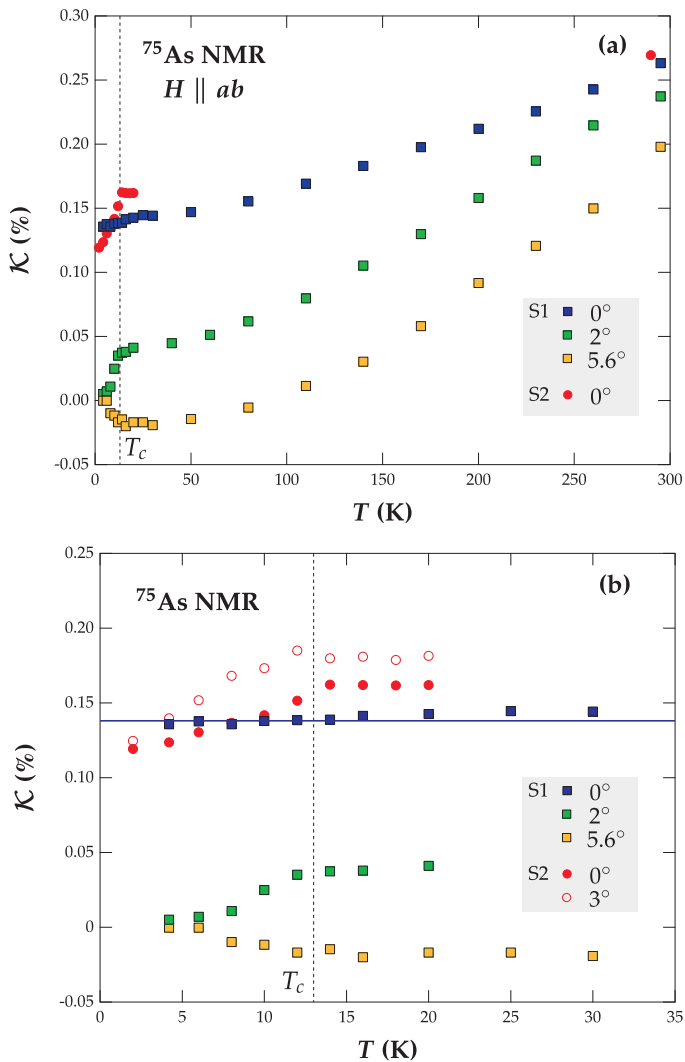}
\caption{\label{fig:knight}(a) The Knight shift ($\mathcal{K}$) as
a function of temperature for sample S1 for different
orientations, and for S2 for $H \parallel ab$. (b) $\mathcal{K}$
at low temperatures. For S1 and $H \parallel ab$, $\mathcal{K}$ is
constant across $T_c$, whereas it decreases below $T_c$ as soon as
the field is tilted out of the $ab$ plane. In contrast, for the
crystal S2, $\mathcal{K}$ drops below $T_c$ even for $H\parallel
ab$, and the angle dependence of $\mathcal{K}$ is not as strong as
that of S1.}
\end{figure}

\subsection{$^{75}$As NMR and NQR linewidth}

Fig. 6 shows the temperature dependence of the full width at half
maximum (FWHM) for the NMR and NQR spectra. For sample S1, FWHM
increases with decreasing temperature, indicating a growing of
magnetic fluctuations. In contrast, FWHM of sample S2 is nearly
temperature independent. Note that for technical reasons we could
obtain only limited high temperature NMR data for S2 (see also
Section \ref{sampleprep}). The $T$ independent NQR FWHM supports
the fact that the NMR FWHM is also $T$ independent, especially
since its absolute value is very small even at low temperatures.

\begin{figure}
\centering
\includegraphics[width=\linewidth]{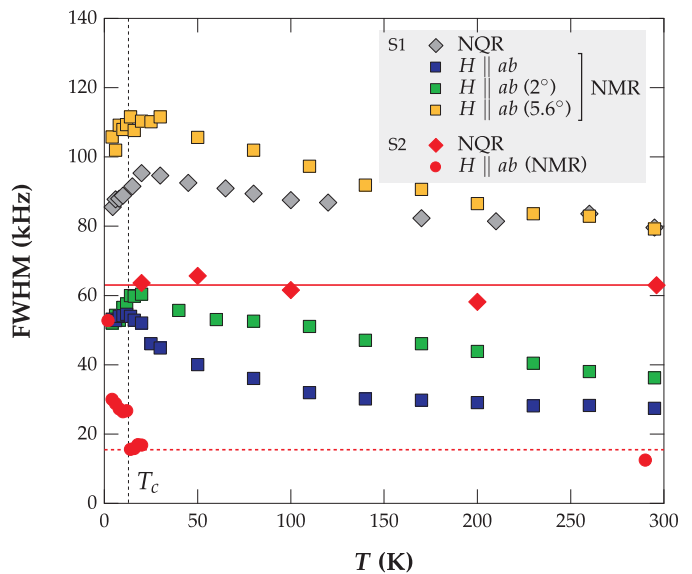}
\caption{\label{fig:FWHM} FWHM as a function of temperature. For
sample S1, FWHM measured by NMR and NQR increases with decreasing
temperature, and shows a strong angle dependence. In contrast, for
the crystal S2, FWHM is smaller than for S1 and the NQR FWHM is
temperature independent. The NMR and NQR FWHM of S2 seem to be $T$
independent as indicated by the dashed line. In the
superconducting state, FWHM for S1 decreases, whereas it increases
for S2.}
\end{figure}

In general, the sample quality is one of the factors determining
the linewidth of NMR and NQR spectra. When interpreting the
linewidth one has to distinguish intrinsic effects such as
broadening from local magnetic moments from the effect of
impurities. This can be done by comparing the temperature
dependence of the linewidth of NMR and NQR spectra where the
broadening from defects (mainly quadrupole effects) and local
magnetic moments (magnetic effects) appear differently. In the
paramagnetic phase, the local moments are fluctuating. If the
inverse correlation time, $1/\tau_c$ of the fluctuations is much
higher than the NMR linewidth, each nucleus sees only the
time-averaged local field and the NMR line should be narrow
(motional narrowing). As the fluctuations slow down, the nuclei
will start to feel a distribution of local fields which broadens
the NMR/NQR lines. Note that even a tiny moment size can broaden
the NMR line considerably. For example, if we assume a moment size
of only $10^{-4} \mu_B$, it will lead to a linewidth on the order
of $\sim 18$ kHz, from the relation $\Delta\nu \sim H_\text{int}
\sim A \mu$, using the hyperfine coupling $A = 6.3$ T/$\mu_B$ as
extracted in Section \ref{knightshift}. The temperature dependence
of FWHM (Fig.~6) corroborates the magnetic broadening. Disorder or
defects in the sample can also lead to an enhanced magnetic
broadening, but the effect of disorder is much stronger on the
quadrupole broadening, and would not lead to the observed
temperature dependence. However, the influence of quadrupole
effects can be directly measured by NQR or by the satellite
transitions ($I_z = -3/2 \leftrightarrow -1/2$ and $I_z = +1/2
\leftrightarrow +3/2$) of the NMR spectrum, whereas the central
transition ($I_z = -1/2 \leftrightarrow +1/2$) is only affected by
second order quadrupole effects. Any small deviations from a
homogeneous charge distribution or lattice anomalies such as
deficiencies or defects will lead to a distribution of EFGs at the
nucleus which contribute to the quadrupole linewidth. Therefore,
while the narrow NQR lines of our LiFeAs single crystals is
related with high homogeneity, the increase of the linewidth when
approaching $T_c$, regardless of the rotation angle, signals that
magnetic correlations progressively gain in strength upon lowering
the temperature in the sample S1, whereas S2 does not show such a
behavior.

We observe that the narrow line for $H \parallel ab$ rapidly
broadens when $H$ rotates out of the plane. A tilting angle of
$2^\circ$ already causes a noticeable broadening (Fig. 6) and for
$5.6^\circ$ the line broadens by a factor of three. For about
$8^\circ$ the width exceeds 200 kHz and $\mathcal{K}$ is very
small. Such an angle dependence could arise from a magnetic
effect, where anisotropic, probably momentum-dependent, spin
fluctuations are present, which are closely related to the
anisotropic Knight shift $\mathcal{K}$. This would be in line with
the presence of incommensurate, nearly ferromagnetic, spin
fluctuations that emerge in microscopic calculations
\cite{brydon11}, the character of which can be strongly affected
by a magnetic field. Or the angle dependence arises from a
quadrupolar effect. In this case, a distribution of angles between
the external magnetic field $H$ and the direction of the principle
axis of the EFG tensor $V_{zz}$ would be needed to explain the
angle dependent broadening, yet narrow NQR resonance line.

In the superconducting state, the linewidth for sample S1
unexpectedly decreases below $T_c$. This feature is also observed
in the NQR spectra. Usually the NMR linewidth increases in the
superconducting state due to vortex-related broadening as is
observed for sample S2. The origin of the decreasing linewidth
just below $T_c$ is not yet clear. Nevertheless, it indicates that
the magnetic properties of the sample change at the
superconducting transition temperature, namely that the spin
fluctuations which lead to the broadening above $T_c$ are strongly
reduced in the superconducting state.

\subsection{$^{75}$As spin-lattice relaxation rates in external field}

By $^{75}$As NQR, we have shown that the $^{75}$As \slr\ is
strongly sample dependent in the superconducting state. In order
to check how an external field affects the low energy spin
dynamics in the superconducting state, we measured \slr\ by
$^{75}$As NMR in the two single crystals S1 and S2. As shown in
Fig. 7, for the crystal S1, we observe that \slr\ even in the
normal state is enhanced compared to the NQR results (solid
horizontal line). Since it is expected that the NMR results for $H
\parallel c$ are equivalent to NQR ones for which the nuclei are
also quantized along the $c$ axis, the behavior is somewhat
surprising, suggesting the presence of unusual, field dependent
low energy spin dynamics in the system. For $H \parallel ab$,
\slr\ is further enhanced, as denoted by the dotted line.
Regardless of the direction of $H$, the strong enhancement of
\slr\ below $T_c$ observed in the NQR measurement (i.e., in zero
field) is substantially suppressed in a magnetic field, although a
sharp drop is still absent for sample S1. A similar behavior has
been found in La$_{0.87}$Ca$_{0.13}$FePO \cite{nakai08}, where
$^{31}$P \slr\ increases just below $T_c$ in low magnetic fields,
while high magnetic fields of about 6 T suppress this increase
(see also the discussion in Section \ref{discuss}).

For the crystal S2, we observe
contrasting results, as in the Knight
shift (Fig.~5). In the normal state \slr\ data for $H \parallel ab$ are
considerably smaller than those of S1. Since NQR results were the
same among all of the samples, this indicates that the spin
dynamics even in the normal state become material-specific in
field. In the superconducting state, a sharp drop at $T_c$ is
observed for S2, as in NQR case for S3.

An interesting observation is that \slr\ of S1 reveals a weak but
clearly visible maximum just above $T_c$ for both directions of
$H$. Together with the enhanced normal state \slr\ in an external
field, one can argue that the normal state spin dynamics are
strongly influenced by an external field, signaling a nearby
critical instability whose nature is unclear yet. Furthermore, the
presence of the local maximum of \slr\ just above $T_c$ in field
as well as the mysterious upturn of \slr\ below $T_c$ in zero
field may indicate unusual superconductivity in LiFeAs.

\begin{figure}
\centering
\includegraphics[width=\linewidth]{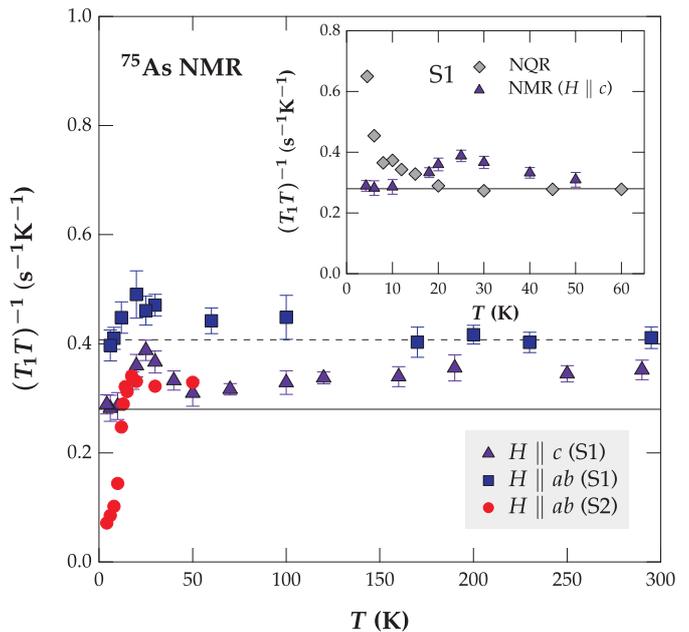}
\caption{$^{75}$As NMR \slr\ as a function of $T$ measured at 7 T
for the sample S1 and at 8.5 T for the sample S2. Above $T_c$ we
observe a small but visible enhancement compared to NQR results
(horizontal solid line), particularly for S1. Below $T_c$, only
the sample S2 exhibits a drop of \slr. Sample S3
exhibits similar temperature dependence as sample S2.}
\end{figure}

\section{Discussion}
\label{discuss}

Our single crystals S1, S2, and S3 of LiFeAs which were grown in
the same environment, exhibit significantly different static
($\nu_Q$ and $\mathcal{K}$) and dynamic [\slr] properties,
particularly, in the superconducting state. Nevertheless, all the
crystals are found to be of good quality and high homogeneity as
evidenced by the narrow $^{75}$As NQR line and the almost equal
linewidth of central and satellite $^{7}$Li NMR lines. Although
the NQR FWHM of S1 ($\sim$80 kHz) is much larger than that
of S3 ($\sim$40 kHz), it is still a factor of more than 2 narrower
than reported for powder LiFeAs samples \cite{li10} and
significantly smaller than reported for other undoped iron
pnictides (see above). Therefore, our results suggest that pure
LiFeAs is located near a critical point so that its physical
properties are extremely sensitive to very small perturbations
such as tiny variations in stoichiometry. A similar critical
doping dependence has also been reported in literature
\cite{pitcher10}, where one percent of Li deficiencies largely
suppress superconductivity. However, we emphasize that all of our
crystals are superconducting with similar $T_c$. We confirmed the onset of 
superconductivity by a strong change in the 
resonance frequency of the NMR sample probe below $T_c$ which is
caused by the change of the surface resistance of the sample.  
Furthermore, a drop of the Knight shift for small angles off $H \parallel ab$ and the 
anomalous change of both the linewidth and \slr\ below $T_c$ indicate that 
sample S1 is also superconducting.

First, we discuss the different behavior of the Knight shift
$\mathcal{K}$ in the superconducting state. When a superconducting
condensate consists of singlet Cooper pairs, a magnetic field
cannot polarize these paired electrons unless pairs are broken up.
Consequently the spin susceptibility $\chi_\text{spin}$ and thus
$\mathcal{K}$ vanishes \cite{yosida58} below $T_c$ when
$T\rightarrow 0$. This may explain the behavior that we observe in
sample S2, and that is reported for powder samples of LiFeAs
\cite{jeglic10,li10} so far. 
In contrast, the constant Knight shift across $T_c$ in sample S1 for 
$H\parallel ab$ exhibits a behavior that would be expected
for a spin-triplet superconductor where pairing does not interfere with the 
magnetic response of the electrons and $\chi_\text{spin}$ remains constant across
$T_c$ down to zero temperature. For small angles off $H \parallel
ab$, however, the Knight shift decreases below $T_c$, suggesting that a
small component of the magnetic field along the $c$ direction
leads to a transition into a state with a superconducting order
parameter of different symmetry in the orbital and/or spin sector.
Note that the so far reported Knight shift measurements of LiFeAs
\cite{jeglic10,li10} have been performed in polycrystalline
samples in which the broadening due to large second order
quadrupole shift may hinder determining the intrinsic spin shift
if there is a strong angle dependence of the Knight shift. Another
explanation would be that the nature of superconductivity may
strongly depend on the effective doping level of the samples, even
though LiFeAs is considered to be a stoichiometric compound. The
differences in the doping level could be visualized by the
slightly different quadrupole frequencies: those samples with a
high quadrupole frequency reveal the constant $\mathcal{K}$ across
$T_c$, whereas those samples with a lower $\nu_Q$, including the
powder sample reported in ref. \cite{li10}, show a more normal
behavior.

A measurement of the intrinsic spin susceptibility in the
superconducting state via the Knight shift is often not straightforward
and other reasons may lead to a constant $\mathcal{K}^{ab}$. First
of all, the correction for the large second order quadrupole shift
could mask a decreasing Knight shift. However, we observe that the
quadrupole frequency exhibits only an insignificant temperature
dependence at low $T$, and the constant $\mathcal{K}^{ab}$ is
already recognizable in the raw data without quadrupole correction
in Fig.~4 (a). Also demagnetization effects are negligible for
$H\parallel ab$ due to the plate-like shape of the crystals.
Furthermore, demagnetization effects reduce the actual field in
the superconductor and therefore should lead to a further decrease
of the total shift $\mathcal{K}$ below $T_c$. A simple heating
effect of the sample by the radio frequency pulses can be excluded
since we observed the decreasing Knight shift for small angles off
$H \parallel ab$ and for sample S2 under similar conditions.
Another reason for a constant $\mathcal K$ in a singlet
superconductor could be an enhanced magnetic susceptibility via
strong orbital magnetism \cite{clogston64} or via spin-orbit
scattering in the presence of disorder \cite{hines71}. Such a
scenario appears rather unlikely. First, the strong angular
dependence of the Knight shift in the superconducting state (see
Fig. 5) is incompatible with a large non-spin contribution to
$\mathcal{K}^{ab}$, such as strong orbital magnetism. For the same
reason the unchanged Knight shift through $T_c$ cannot be
explained by an impurity-based scenario. Moreover, our samples
appear to be clean single crystals, even though tiny differences
between the crystals have to exist.


In LiFeAs the ($\pi$,$\pi$)-nesting and static antiferromagnetism
are absent \cite{borisenko10}. Investigating theoretically the
magnetic and pairing instabilities in an electronic model that
incorporates the poor nesting properties and unusually shallow
hole pockets of LiFeAs, Brydon and coworkers \cite{brydon11} find
ferromagnetic fluctuations to drive an instability toward
spin-triplet $p$-wave superconductivity. These magnetic
fluctuations are related to LiFeAs being in the vicinity of
nearly ferromagnetic, incommensurate, long-range
spin ordered phases, the stability of which is governed by the
detailed electronic density and interaction parameters. 
Such a spin-triplet scenario may be supported by the constant spin 
susceptibility across $T_c$ observed for $H\parallel ab$ and the increasing 
linewidth with decreasing temperature in sample S1, although 
samples S2 and S3 clearly show a spin-singlet behavior.  
As we have discussed above,
the experimental results in single crystals S1--S3 should result
from intrinsic properties rather than extrinsic ones. This
suggests that different phases are competing in LiFeAs where
relatively small changes due to external fields, structure or
stoichiometry can cause transitions between unconventional and
more conventional superconducting ground states, and may also
explain other contradicting experimental results in literature
(see also Section \ref{intro}).

Another anomalous behavior is the strange upturn of \slr\ in sample S1 and in 
the polycrystalline sample. Nakai et al. \cite{nakai08} suggest four
different possibilities for a similar upturn in
La$_{0.87}$Ca$_{0.13}$FePO: (i) impurity contributions which can
be excluded for both compounds, as argued above. Furthermore, the
relaxation curves are single exponential in LiFeAs, too, whereas
impurities would lead to a multiexponential behavior of the
relaxation curves. (ii) Vortex contributions to \slr\ which can be
excluded for LiFeAs since the upturn occurs in zero magnetic field
(NQR). (iii) Slowing of magnetic fluctuations due to the opening
of the superconducting gap. In this case, scattering of local
magnetic moments with conduction electrons in the normal state
enhances the energy scale of the fluctuations. In the
superconducting state the scattering is suppressed due to the
formation of Cooper pairs, and the local magnetic moments may slow
down. As argued by Nakai et al., such a behavior has not been
reported yet. (iv) Collective modes of the spin-triplet pairs
could give rise to novel spin dynamics in the SC state
\cite{nakai08,vollhardt}. In particular, we note that the possibility (iv) may 
explain the observed constant Knight shift in the very same sample S1.

In order to reconcile the two different results even in the single
crystals grown under the same conditions, we conjecture that the
nature of spin fluctuations, which is generally thought to mediate
Cooper pairs in non-BCS unconventional superconductors, could be
either antiferromagnetic or ferromagnetic depending on the
proximity to a ferromagnetic instability in an extremely sensitive
fashion. 
It is also interesting to note that there is a seeming trend
that the EFG and thereby the NQR frequency $\nu_Q$ are larger, when
the Knight shift and \slr\ exhibit the peculiar behaviors. This
indicates a possible connection between the doping level, which
could be proportional to the EFG as in other iron pnictides, and a
ferromagnetic instability. Consistently, Co doping leads to a
lower quadrupole frequency (EFG), and to a reduction of \slr\ in
the superconducting state, putting the Co doped sample further
away from the ferromagnetic instability. This observation is also
consistent with theory \cite{platt11,brydon11}, where the small
hole pocket at the $\Gamma$ point drives the ferromagnetic
fluctuations. Regarding Co doping as electron doping, the hole
pocket should shrink upon doping, thereby weakening the
ferromagnetic fluctuations \cite{aswartham11}.

\section{Conclusion}

We have used NMR and NQR to investigate different LiFeAs samples.
We find that tiny differences in the stoichiometry of the samples
exist which lead to different normal state properties as well as
to different superconducting properties. A possibility to
distinguish the samples is the quadrupole frequency, $\nu_Q$.
Sample S1 yields a slightly higher $\nu_Q$ than sample S2, whereas
$\nu_Q$ of a Co doped sample is even below that of S2. At the same
time, the Knight shift of sample S1 is constant across $T_c$ for
$H \parallel ab$. Although this may suggest spin triplet superconductivity, 
we find that the Knight
shift drops below $T_c$ for small angles off the $ab$ plane. 
On the other hand the
Knight shift of sample S2 decreases in the SC state regardless of 
the angle which is compatible with standard singlet pairing of the Cooper
pairs. The temperature dependence of the linewidth varies also 
with the crystals.  The linewidth of sample S1 increases with decreasing 
temperature, indicating a growing of magnetic fluctuations, while
that of sample S2 is temperature independent. Consistent
with the observation of $\mathcal{K}^{ab}=\text{ const.}$ across
$T_c$ could be the increasing \slr\ in the SC state
of sample S1 as well as of the polycrystalline sample which also
yields a higher $\nu_Q$. The increase of \slr\ below $T_c$
could originate from non-vanishing spin degree of freedom in the SC state which
give rise to novel spin dynamics.

\section*{Acknowledgement}

We are deeply grateful to K. Kitagawa and M. Takigawa for
invaluable experimental collaboration and for sharing their data
with us. The authors also thank G. Lang, C. Nacke, I. Morozov, M.
Daghofer, C. Timm, and P. M. Brydon for discussion and M.
Deutschmann, J. Werner, A.  Voss, J.  Eckert, and R. Vogel for
technical support. This work has been supported by the Deutsche
Forschungsgemeinschaft through FOR 538 (Grant No. BU887/4) and
SPP1458 (Grant No. GR3330/2 and BE1749/13). SW acknowledges
support by DFG under the Emmy-Noether program (Grant No.
WU595/3-1).

\bibliography{mybib}

\end{document}